# Generative AI-Driven Decision-Making for Disease Control and Pandemic Preparedness Model 4.0 in Rural Communities of Bangladesh: Management Informatics Approach

**Mohammad Saddam Hosen** ✉

M.Phil (Researcher), Department of Management Studies, National University, Gazipur-1704, Bangladesh

**MD Shahidul Islam Fakir**

Professor, Department of Management Studies, Jagannath University, Bangladesh

**Shamal Chandra Hawlader**

Deputy Director (Deputation), Rural Development Academy (RDA), Bangladesh

**Dr. Farzana Rahman**

Medical Officer, Department of Psychiatry, Mugda Medical College, Bangladesh

**Dr. Tasmim Karim**

Senior Lecturer, Department of Anatomy, Pioneer Dental College, Bangladesh

**Muhammed Habil Uddin**

Lecturer, Information & Communication Technology, Dhaka Public College, Bangladesh

## Abstract

Rural Bangladesh is confronted with substantial healthcare obstacles, such as inadequate infrastructure, inadequate information systems, and restricted access to medical personnel. These obstacles impede effective disease control and pandemic preparedness. This investigation employs a structured methodology to develop and analyze numerous plausible scenarios systematically. A purposive sampling strategy was implemented, which involved the administration of a questionnaire survey to 264 rural residents in the Rangamati district of Bangladesh and the completion of a distinct questionnaire by 103 healthcare and medical personnel. The impact and effectiveness of the study are assessed through logistic regression analysis and a pre-post comparison that employs the Wilcoxon Signed-Rank test and Kendall's coefficient for non-parametric paired and categorical variables. This analysis evaluates the evolution of disease control and preparedness prior to and subsequent to the implementation of the Generative AI-Based Model 4.0. The results indicate that trust in AI ($\beta$ = 1.20, p = 0.020) and confidence in sharing health data ($\beta$ = 9.049, p = 0.020) are the most significant predictors of AI adoption. At the same time, infrastructure limitations and digital access constraints continue to be significant constraints. The study concludes that the health resilience and pandemic preparedness of marginalized rural populations can be improved through AI-driven, localized disease control strategies. The integration of Generative AI into rural healthcare systems offers a transformative opportunity, but it is contingent upon active community engagement, enhanced digital literacy, and strong government involvement.

### Introduction

Multifaceted healthcare challenges in rural Bangladesh are the result of inadequate information systems, limited access to medical professionals, and a lack of infrastructure. Poor health outcomes are frequently the result of underserved areas [1], particularly during epidemics and pandemics. In remote areas, over 70% of the rural population has limited access to primary healthcare [2]. The COVID-19 pandemic has emphasized the necessity of rapid response systems and pandemic preparedness [3]. Rural healthcare systems were found to be inadequately prepared as a result of insufficient data [4], limited resources, and a lack of coordinated information-sharing mechanisms [5]. Bangladesh has reported over 2 million COVID-19 cases and 29,434 fatalities despite efforts to control the spread through vaccination campaigns [6], which are still insufficient, particularly in light of emerging variants [7]. The potential to enhance disease control and preparedness is demonstrated by the integration of Generative Artificial Intelligence (GAI) and data-driven decision-making into healthcare [8], [9], [10], [11]. Management Informatics Networks (MIN) that integrated the GAI system [12], and effectively predicted the spread of disease, simulated epidemic scenarios, and recommended interventions in urban areas [13].

Nevertheless, their application in resource-constrained rural environments such as Bangladesh is still largely unexplored. The absence of dependable data for decision-making is a significant obstacle, as the availability of digital health records and real-time data collection is restricted, which impedes the development of effective responses [14]. In order to construct precise disease simulations, identify at-risk areas, and plan targeted pandemic strategies for rural Bangladesh, GAI can fill these gaps by synthesizing data from a variety of sources [15]. Pillai and Pillai [16] have demonstrated the efficacy of AI in predicting disease outbreaks and facilitating pandemic preparedness in rural areas [17]. For example, the utilization of localized data by AI models in rural India has resulted in enhanced outbreak predictions [7], [18]. This research demonstrates the potential of AI-driven solutions to improve pandemic preparedness in rural Bangladesh. It proposes the Generative AI-Based Disease Control and Pandemic Preparedness Model 4.0 to enhance healthcare responsiveness in these communities [19]. The following research questions were used to investigate the incorporation of Generative AI in enhancing disease control and pandemic preparedness in rural Bangladesh:

▪ In rural Bangladesh, how can the accuracy of disease outbreak predictions and preparedness strategies be enhanced through the use of Generative AI-based models?

▪ What are the primary obstacles to the implementation of management Informatics-based GAI systems in rural healthcare infrastructures in Bangladesh?

▪ How can the efficacy of AI-driven disease control models 4.0 in rural Bangladesh be improved through localized data collection and real-time surveillance initiatives?

The potential impacts, challenges, and feasibility of implementing Generative AI in rural healthcare systems in Bangladesh were elucidated by addressing these inquiries, with the objective of improving disease control and pandemic preparedness [20]. GAI in rural Bangladeshi healthcare systems can improve disease prediction, preparedness, and policy creation. GAI models predict disease outbreaks using complicated health data, enabling prompt interventions [21], [22]. AI can also identify implementation hurdles like infrastructural issues and digital literacy issues and use localized data collecting to tailor disease management models to remote communities.

The public health authorities of Bangladesh have significantly enhanced the preparedness and control measures by model 4.0 for a variety of epidemics [23], [24] such as the Dengue Fever epidemic [25], Cholera outbreak, H1N1 Influenza (Swine Flu), Typhoid fever epidemic, Avian Influenza (Bird Flu), Hepatitis E outbreak, Measles outbreak, Malaria outbreak, Polio, Zika and Ebola virus, MERS-CoV, Asian Flu, and Norovirus outbreaks [26]. This has ultimately aided in the protection of rural communities. That data-driven decision-making approach in model 4.0 could assess epidemic patterns to help avert outbreaks [27], [28], and improve rural healthcare in Bangladesh by tackling these obstacles and using AI to promote health equity and resilience in marginalized regions.

This review of the literature investigates the integration of Generative AI in pandemic preparedness and disease control, with a particular emphasis on its implementation in rural healthcare settings, particularly in developing countries. The review is structured as shown in Table 1:

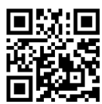

Table 1: Literature Review Findings

| Sources | Methods | Publications Type | Findings |
|---|---|---|---|
| Tariq [29] | Risk Prediction | Book Chapter | Predictive modelling with GAI may improve pandemic preparedness and preemptive actions. Data-driven disease control decisions are crucial in rural Bangladesh. |
| Ko and Ogiela [30] | Blockchain and GAI Framework | Journal Article | Blockchain increases digital medical content security in AI-driven healthcare systems, says the study. It emphasises decentralised medical data privacy and dependability security. |
| Bosco et al. [31] | Apps-based Usability Study | Journal Article | It examines the design and use of a multimodal AI tool for Black American informal caregivers managing Alzheimer's and dementia. It emphasizes cultural relevance and user-centered design to improve AI-driven healthcare products' accessibility and efficacy. |
| Lechien [32] | GAI on Otolaryngology | Journal Article | GAI improves diagnoses, surgical planning, and patient care in otolaryngology, according to the study. It shows that AI-driven models can improve head and neck surgery clinical decision-making and tailored treatment. |
| Khamparia and Gupta [33] | Augmentation techniques | Book | For predictive modeling and personalized healthcare, the book examines how generative AI has transformed biomedical research and smart health informatics. Data analysis with AI improves disease management, diagnosis, and healthcare decisions. |
| Sai et al. [9] | GAI Predictive Analytics | Conference Paper | The study examines healthcare generative AI models' applications, case studies, and limits. It shows how AI-driven decision-making can improve disease control, patient management, and healthcare efficiency. |
| Ray [34] | GAI Model, Quantitative Approach | Journal Article | GAI is changing metabolic dysfunction-associated fatty liver disease research and treatment, according to the study. It shows how AI-driven predictive models improve early diagnosis, individualized treatment, and patient outcomes. |
| Ali et al. [35] | Deep GAI Models | Book | This book examines recent advances in deep generative models and medical AI applications to improve diagnosis and therapy. It shows how AI may improve data-driven decision-making for more accurate and efficient healthcare solutions. |
| Moulaei et al. [36] | PRISMA-ScR based Scoping review | Journal Article | This scoping review discusses the benefits, drawbacks, and many uses of generative AI in healthcare, focusing on patient care and medical decision-making. AI-driven models improve disease prediction, diagnosis, and individualized treatment while resolving ethical and implementation issues. |
| Albaroudi et al. [37] | GAI Trained Technique | Conference Paper | In this study, generative AI is used to overcome significant obstacles in patient care and medical decision-making. AI improves diagnostic accuracy, treatment customization, and healthcare efficiency. |
| Letafati and Otoum [38] | AI Database Case Study | Magazine | This study addresses metaverse-integrated digital healthcare and privacy and security. It emphasizes the necessity for strong data protection and regulatory frameworks to ensure safe and ethical virtual healthcare delivery. |
| Params [39] | AI Clinical Diagnosis, Image Analysis | Journal Editorial | Artificial intelligence in infectious disease surveillance is important for epidemic and pandemic preparedness, according to this editorial. AI-driven models can improve |

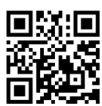

| | | | early detection, real-time monitoring, and data-driven decision-making to stop infectious disease spread. |
|---|---|---|---|
| Kwok et al. [40] | GAI Model Developed | Journal Article | To improve public health preparedness, this study examines infectious disease transmission modeling using large language models (LLMs). It shows how AI-driven predictive analytics help policymakers make timely and effective disease management decisions. |

This expanded Table 1 further elaborates on the existing literature, with a particular emphasis on pandemic preparedness, data-driven decision-making, and AI, particularly in the context of rural Bangladesh. It is indicative of the ways in which a variety of studies serve to enhance comprehension of the technological solutions and obstacles associated with the management of public health crises in rural areas.

**Materials and Methods**

The study enhances its profundity and relevance by management informatics employing systematic scenario modeling [41], [42], which enables the examination of potential outcomes and broader implications. A purposive sampling approach was implemented [43], which involved the administration of a Likert-scale questionnaire to 264 rural residents from the Rangamati district of Bangladesh. These residents were drawn from the riverside, plainland, and Chittagong Hill Tracts (CHT) communities. Furthermore, a distinct questionnaire was completed by 103 healthcare personnel. The data that was gathered was instrumental in the creation of a disease control model that was AI-driven and customized.

The AI model's effectiveness and adoption were evaluated through a pre-post comparison that utilized Wilcoxon Signed-Rank and Kendall's tests for non-parametric paired and categorical variables, as well as logistic regression. SPSS v27 was employed to conduct the statistical analysis, and Microsoft Excel 2019 was employed to generate visual representations of the results. The study also assessed the applicability of Generative AI Model 4.0 in rural Bangladeshi healthcare systems.

Data acquisition from a variety of rural communities was the initial step in the structured process that the research followed. Subsequently, the data was incorporated into the Generative Artificial Intelligence Model 4.0 (GAIM 4.0), and a comparative analysis was conducted to evaluate the differences among communities, as illustrated in Figure 1. In order to enhance decision-making in pandemic preparedness and disease control, scenario planning was implemented to simulate various health conditions. The methodology was designed to meet the unique requirements of rural populations through AI-driven predictive modeling and scenario analysis.

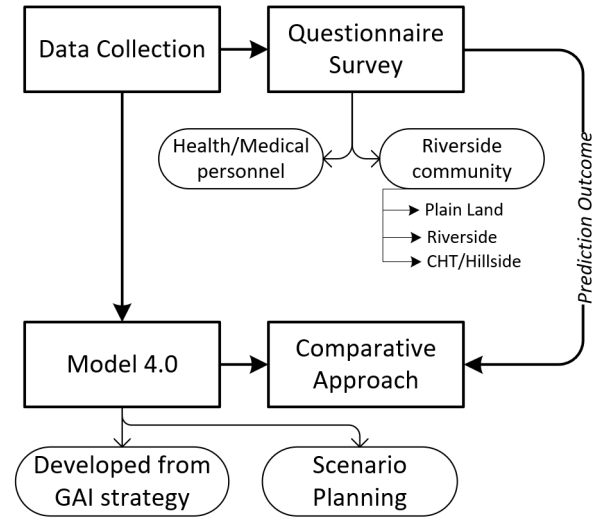

**Figure 1: Study Method Design**

**Results and Discussion**
**GAI-Based Model 4.0**
Several sophisticated components have been incorporated into the GAI-Based Disease Control and Pandemic Preparedness Model 4.0 to enhance disease control and pandemic readiness in Figure 2, particularly in resource-limited environments.

*Pandemic and Disease Data Collection*
The model commences with the critical Pandemic and Disease Data Collection phase, which entails the acquisition of data in a variety of forms to monitor disease outbreaks. Examples of these include syndromic surveillance, mobile health (mHealth) monitoring, case and contact tracing, and surveys. Each data collection method is crucial for the collection of real-time and historical health data, which is necessary for the analysis of community health status and the monitoring of disease trends [44]. Data gathered through these mechanisms serves as the basis for subsequent analysis and judgment.

*Statistical and Historical Analysis: Pandemic, Outbreak, and Virus*
The pandemic's progression, previous outbreaks, and the virus itself are analyzed through Statistical and Historical Analysis following data accumulation. Descriptive analytics (to summarize data), predictive analytics (to forecast future trends), prescriptive analytics (to recommend optimal actions), and diagnostic analytics (to identify the underlying causes of outbreaks) are all included in this step. Furthermore, geospatial and sentiment analysis enhances the model's accuracy by providing geographical context and gauging public sentiment regarding health measures [45]. The objective of this phase is to evaluate the efficacy of existing strategies, identify disease patterns, and offer insights for opportune interventions in Figure 2.

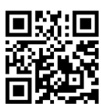

*GAI Data Analysis: Data Augmentation and Synthesis*

The subsequent phase employs Generative AI Data Analysis, which enhances the predictive capabilities of the model by utilizing machine learning and AI algorithms to process large datasets. AI enables the generation of new data from existing datasets, thereby enhancing model performance in cases where data may be incomplete through Data Augmentation and Synthesis. Another critical function of this phase is pandemic forecasting, which employs artificial intelligence to simulate and foresee potential future outbreaks (Figure 2). By analyzing trends and determining optimal containment strategies, AI-powered systems also contribute to Improved epidemic modeling [46].

Additionally, Data Imputation ensures that the decision-making process is founded on comprehensive, accurate data by addressing data gaps in incomplete datasets. Parallel Data Transformation, which encompasses Natural Language Processing (NLP) and Natural Language Generation (NLG), further improves public health surveillance by extracting and organizing unstructured data, such as news reports or social media posts, into actionable health insights [47].

*Enhancing the Decision Support System*

The Decision Support System (DSS) is a critical component of the model [48], as it consolidates data analysis from the AI system and presents it in a format that is actionable for healthcare professionals and policymakers. This system guarantees informed decision-making at the local and national levels by synthesizing predictions, risk assessments, and resource recommendations. The system facilitates dynamic modifications to health strategies by providing continuous feedback and updates in response to new data.

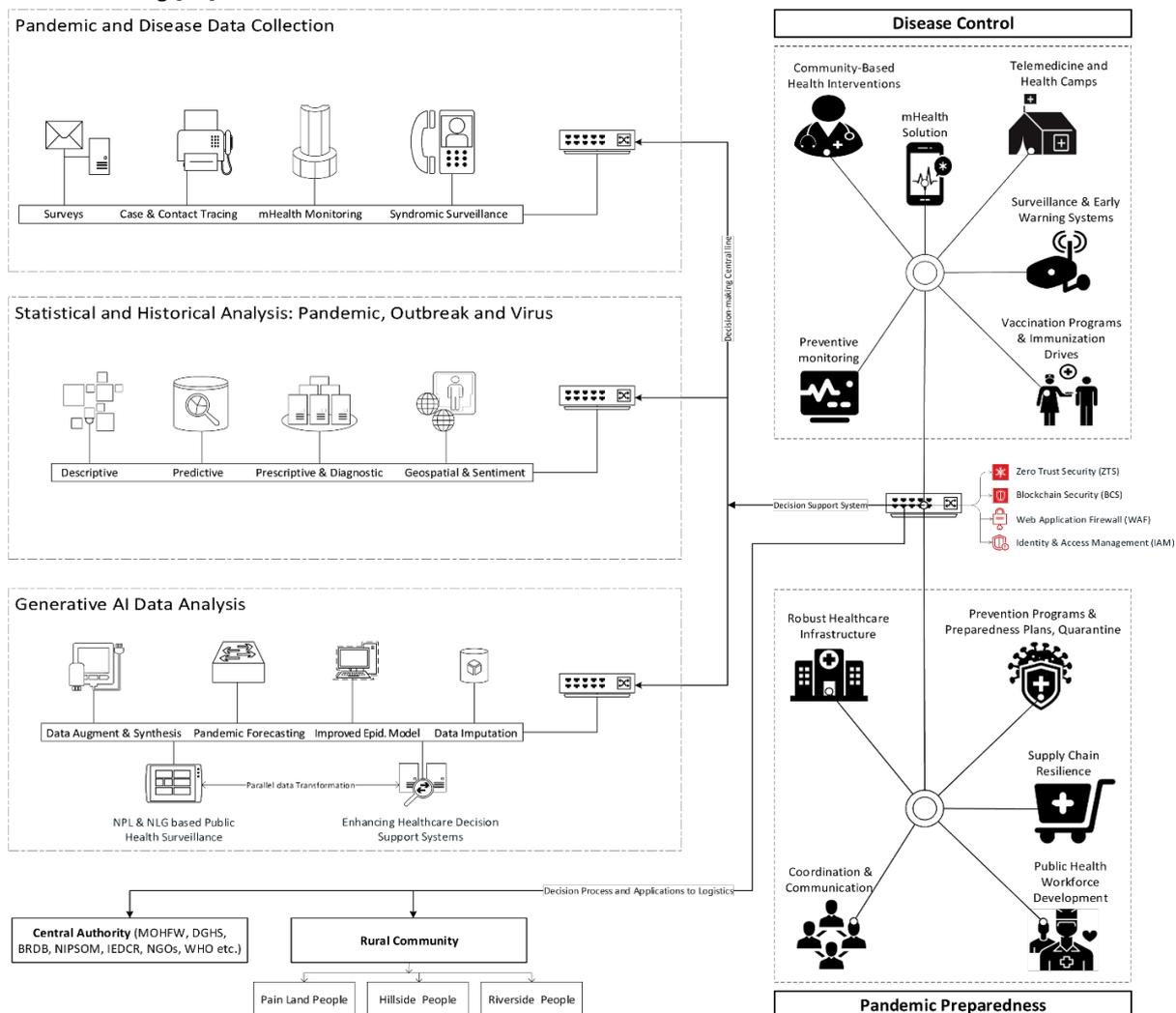

**Figure 2: GAI-Based Disease Control and Pandemic Preparedness Model 4.0**

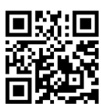

*Central Authority and Rural Community Interaction*
The model recognizes the critical role of Central Authorities (e.g., Ministry of Health and Family Welfare, Directorate General of Health Services, non-governmental organizations, and other public health entities) in directing the pandemic response. The central authorities are accountable for the coordination of resources, the implementation of strategies [49], and the supervision of the overall management of disease control initiatives.

Conversely, the health interventions are directed toward the Rural Community, which encompasses the Plain Land, Hillside, and Riverside populations. The model underscores the necessity of context-specific, localized strategies, which are designed to address the distinctive obstacles encountered by rural communities (Figure 2), including inadequate infrastructure and healthcare access [50]. Localized responses to disease hazards are made possible by the data that these communities provide to the DSS.

*Disease Control and Pandemic Preparedness*
A variety of intervention strategies, such as Community-Based Health Interventions, Telemedicine, and Vaccination Programs, are implemented during the Disease Control phase (Figure 2). These interventions are designed to mitigate the transmission of diseases, alleviate current epidemics, and anticipate potential hazards in the future. In rural and marginalized regions [51], the timely detection and response allowed by the use of Surveillance and Early Warning Systems are critically important. These systems notify authorities of potential disease outbreaks, which in turn facilitate the implementation of preventive measures.

The model's primary objective during the Pandemic Preparedness stage is to guarantee the readiness of healthcare systems and infrastructures [52]. This entails the enhancement of the Public Health Workforce through training and resources, the reinforcement of Coordination and Communication mechanisms between local communities and national authorities, and the reinforcement of Supply Chain Resilience to ensure continuous access to medical supplies.

*Security Measures*
The data and the integrity of the decision-making process are safeguarded by the model's robust security measures. These systems comprise Blockchain Security (BCS), Web Application Firewalls (WAF), Zero Trust Security (ZTS), and Identity & Access Management (IAM). These security protocols guarantee that the data collected and analyzed by the DSS is protected from external threats, thereby protecting the health data of the public and the functionality of the AI-driven systems focused in Figure 2. The utilization of blockchain technology enhances the transparency and immutability of health records, while Identity and Access Management (IAM) systems safeguard sensitive information by limiting access to authorized personnel [53].

The GAI-Based Disease Control and Pandemic Preparedness Model 4.0 is a sophisticated framework that employs data analytics and AI technologies to improve the response to disease in rural areas. To aid both central authorities and local communities in the management of health crises, it integrates real-time data, historical analysis, predictive modeling, and decision support. The model offers a comprehensive strategy for addressing future health hazards and enhancing pandemic resilience, incorporating AI tools, security measures, and preparedness strategies.

**Effectiveness and Impact Measures**
In this study, the efficacy of AI-based disease control models in rural Bangladesh is assessed, with an emphasis on the factors of familiarity with AI, confidence in its predictive capabilities, and barriers to adoption. The reliability evaluations of the measurement model guarantee its consistency, while logistic regression offers a thorough examination of the perspectives of health personnel. Here is the logistic regression equation [54]:

$$Logit(p) = In\left(\frac{p}{1-p}\right) \quad (1)$$

$$= \beta_0 + \beta_1 \,(Familiarity\ with\ GAI) + \cdots \beta_7 (Other\ predeictors)$$

The study also implements comparative evaluations to assess the distinctions between groups. The Wilcoxon Signed Rank Test is implemented on paired observations, employing the following equation [55]:

$$W = \min(\sum(ranks\ for\ positive\ differences) \\ \sum(ranks\ for\ negative\ differences))$$

In addition, Kendall's Coefficient of Concordance is employed to evaluate the extent of consensus regarding the efficacy of AI in disease control, as denoted by [56]:

$$W = \frac{12 \sum R_i^2 - 3n\,(n+1)^2}{n^2(k^3 - k)} \quad (2)$$

The success of AI in disease control is predicted by regression models, while the relationship between trust in AI and its perceived efficacy is investigated by correlation analysis. In general, the results underscore the significance of government involvement and identify the primary barriers and facilitators to the adoption of AI in rural healthcare.

*Medical Personnel Evaluation*
Health and medical personnel' views on healthcare issues and AI-driven solutions in rural Bangladesh are examined in this section. It analyzes how the GAIM 4.0 for disease control and pandemic preparedness improves healthcare resilience and preparedness through data-driven decision-making.

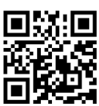

**Table 2: Descriptive Statistics by Health and Medical Personnel**

| Variables (N=103) | Mean | SEM | SD | Var. | Skewness | Kurtosis |
|---|---|---|---|---|---|---|
| 1. Gender distribution of respondent. | 1.243 | 0.04 | 0.431 | 0.19 | 1.22 | -0.53 |
| 2. The age group of respondents. | 2.786 | 0.14 | 1.405 | 1.97 | 0.58 | -0.56 |
| 3. Profession. | 4.922 | 0.41 | 4.137 | 17.11 | 0.74 | -0.82 |
| 4. The concept of GAI (Generative Artificial Intelligence) in healthcare is familiar to many people. | 3.534 | 0.14 | 1.399 | 1.96 | -0.48 | -0.87 |
| 5. AI-based models have the potential to improve disease prediction and pandemic preparedness in rural communities. | 3.107 | 0.11 | 1.137 | 1.29 | 0.15 | -1.11 |
| 6. AI-based models are considered reliable in predicting disease outbreaks in rural areas. | 2.903 | 0.15 | 1.537 | 2.36 | 0.05 | -1.47 |
| 7. Access to healthcare services in rural communities is often challenged by various factors. | 2.136 | 0.12 | 1.205 | 1.45 | 0.93 | 0.15 |
| 8. Many people are comfortable sharing personal health data (e.g., symptoms and medical history) with AI-driven systems to help predict disease outbreaks. | 3.476 | 0.17 | 1.691 | 2.86 | -0.42 | -1.56 |
| 9. Medical history, demographics, and symptoms are key data for AI-based disease control systems. | 2.417 | 0.10 | 1.024 | 1.05 | 1.15 | 0.61 |
| 10. The adoption of new technology, such as AI-based disease control systems, is likely to occur when healthcare access in rural communities is improved. | 3.282 | 0.13 | 1.331 | 1.77 | -0.15 | -0.91 |
| 11. Real-time surveillance systems, like mobile health monitoring, could help in the early detection of disease outbreaks. | 3.272 | 0.13 | 1.277 | 1.63 | -0.70 | -0.77 |
| 12. Several barriers exist to implementing AI-driven disease control models in rural Bangladesh. | 3.612 | 0.14 | 1.443 | 2.08 | -0.77 | -0.70 |
| 13. Involving rural communities in the development and deployment of GAI-based disease control systems is crucial for government and health authorities. | 3.282 | 0.13 | 1.331 | 1.77 | -0.15 | -0.91 |

**Note:** SEM = Standard Error of Means, SD = Standard Deviation, Var. = Variance

**Table 3: Frequency Distribution by Health and Medical Personnel**

| Variable | Variable Categories | Count (N=103) | Percent |
|---|---|---|---|
| Sex | Male | 78 | (75.73%) |
|  | Female | 25 | (24.27%) |
| Age group | 18 to 24 | 19 | (18.45%) |
|  | 25 to 34 | 33 | (32.04%) |
|  | 35 to 44 | 22 | (21.36%) |
|  | 45 to 54 | 13 | (12.62%) |
|  | 55 to 64 | 12 | (11.65%) |
|  | 65 or over | 4 | (3.88%) |
| Profession | Doctor | 30 | (29.13%) |
|  | Nurse | 17 | (16.50%) |
|  | Medical Technologist | 7 | (6.80%) |
|  | Pharmacist | 5 | (4.85%) |
|  | Dentist | 6 | (5.83%) |
|  | Medical Assistant | 3 | (2.91%) |
|  | Physiotherapist | 4 | (3.88%) |

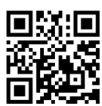

| | | | |
|---|---|---|---|
| | Dietitian/Nutritionist | 6 | (5.83%) |
| | Teacher (Public Health) | 6 | (5.83%) |
| | Epidemiologist | 2 | (1.94%) |
| | Public Health Specialist | 10 | (9.71%) |
| | Disease Surveillance Officer | 1 | (0.97%) |
| | Infectious Disease Specialist | 2 | (1.94%) |
| | Health Program Coordinator | 4 | (3.88%) |
| GAI concept in healthcare | Very unfamiliar | 15 | (14.56%) |
| | Somewhat unfamiliar | 2 | (1.94%) |
| | Neutral | 39 | (37.86%) |
| | Somewhat familiar | 7 | (6.80%) |
| | Very familiar | 40 | (38.83%) |
| GAI models improve prediction | Strongly disagree | 4 | (3.88%) |
| | Disagree | 36 | (34.95%) |
| | Neutral | 21 | (20.39%) |
| | Agree | 29 | (28.16%) |
| | Strongly agree | 13 | (12.62%) |
| GAI models predict outbreaks | Not reliable at all | 30 | (29.13%) |
| | Slightly unreliable | 13 | (12.62%) |
| | Neutral | 20 | (19.42%) |
| | Slightly reliable | 17 | (16.50%) |
| | Very reliable | 23 | (22.33%) |
| Healthcare access challenges rural | Lack of healthcare facilities | 41 | (39.81%) |
| | Limited access to medical professionals | 26 | (25.24%) |
| | Poor infrastructure (roads, transportation) | 25 | (24.27%) |
| | High cost of healthcare | 3 | (2.91%) |
| | Lack of technology or internet access | 8 | (7.77%) |
| | Other | 0 | (0.00%) |
| Comfortable sharing health data | Very uncomfortable | 23 | (22.33%) |
| | Uncomfortable | 12 | (11.65%) |
| | Neutral | 13 | (12.62%) |
| | Comfortable | 3 | (2.91%) |
| | Very comfortable | 52 | (50.49%) |
| Useful data: history, demographics, symptoms | Historical health data (e.g., past outbreaks) | 10 | (9.71%) |
| | Real-time health data (e.g., symptoms, vaccination rates) | 65 | (63.11%) |
| | Environmental data (e.g., weather, water quality) | 9 | (8.74%) |
| | Social and demographic data (e.g., age, gender, population density) | 13 | (12.62%) |
| | Other | 6 | (5.83%) |
| Adoption of GAI improves access | Very unlikely | 14 | (13.59%) |
| | Unlikely | 9 | (8.74%) |
| | Neutral | 43 | (41.75%) |
| | Likely | 8 | (7.77%) |
| | Very likely | 29 | (28.16%) |
| Real-time surveillance detects outbreaks | Strongly disagree | 17 | (16.50%) |
| | Disagree | 11 | (10.68%) |
| | Neutral | 13 | (12.62%) |
| | Agree | 51 | (49.51%) |
| | Strongly agree | 11 | (10.68%) |
| Barriers to GAI Implementation in rural | Lack of internet access | 18 | (17.48%) |
| | Limited understanding of AI among healthcare workers | 2 | (1.94%) |
| | Resistance to new technology from the community | 20 | (19.42%) |

| | Financial constraints | 25 | (24.27%) |
|---|---|---|---|
| | Lack of trained professionals in AI and technology | 38 | (36.89%) |
| | Other | 0 | (0.00%) |
| Health professional involvement in GAI development | Not important at all | 14 | (13.59%) |
| | Slightly important | 9 | (8.74%) |
| | Neutral | 43 | (41.75%) |
| | Important | 8 | (7.77%) |
| | Very important | 29 | (28.16%) |

The descriptive statistics in Table 2 for the 103 respondents indicate that the data reflects a variety of opinions and perceptions regarding AI-driven disease control systems in rural Bangladesh. The mean values indicate a general understanding of Generative Artificial Intelligence (GAI) and its potential advantages. The mean values for queries regarding the sharing of personal health data (M = 3.476) and the involvement of rural communities in the development of GAI-based systems (M = 3.282) are higher. The data also suggests that the concept that AI-based models can enhance disease prediction and pandemic preparedness is significantly supported, with a mean of 3.107. The skewness values primarily indicate a minor positive skew, suggesting a tendency toward agreement with the statements, despite the fact that some variability exists across the responses, as evidenced by the standard deviations (ranging from 1.137 to 1.691). For instance, the statement regarding the reliability of AI-based models in predicting disease outbreaks exhibited a positive deviation (0.58), indicating that a greater number of respondents were inclined to concur. The kurtosis values indicate platykurtic distributions (negative kurtosis), which implies that the data does not exhibit heavy tails or extreme anomalies but rather follows a more uniform distribution across the responses. The respondents' moderate support and comprehension of the potential integration of AI in enhancing healthcare outcomes in rural Bangladesh are suggested by these results.

The sample is predominantly composed of male participants (75.73%), with the remaining 24.27% being female, as indicated by the frequency distribution of the health and medical personnel's responses in Table 3. The age group of 25 to 34 years old comprises the plurality of respondents (32.04%), with the 35 to 44 age group following closely behind (21.36%). The profession with the highest percentage of individuals is physicians (29.13%), followed by nurses (16.50%) and medical technologists (6.80%). The majority of respondents (38.83%) reported that they were extremely familiar with the concept of Generative Artificial Intelligence (GAI) in healthcare, while 37.86% were neutral, suggesting a degree of understanding. The majority of participants, 40.77%, concur that GAI models have the potential to enhance disease prediction and pandemic preparedness. However, 34.95% of the responses were neutral regarding the reliability of these models in predicting outbreaks. The accessibility of healthcare services in rural communities was identified as a significant challenge, with the most prevalent obstacles being a lack of healthcare facilities (39.81%) and limited access to medical professionals (25.24%). In terms of the sharing of health data, the majority of participants (50.49%) expressed a high level of familiarity with the idea of sharing personal health information with AI-driven systems. Additionally, 63.11% of participants viewed real-time health data (including vaccination rates and symptoms) as the most beneficial for AI disease control systems. The adoption of GAI to enhance healthcare access in rural areas is perceived as probable by 41.75%. However, the lack of trained professionals in AI and technology (36.89%) and financial constraints (24.27%) were identified as substantial obstacles to GAI implementation. Finally, 28.16% of health professionals deemed their participation in the development of GAI-based disease control systems to be of critical importance, suggesting a strong desire to contribute to this endeavor.

**Table 4: Likelihood Ratio**

| Effect | Model Fitting Criteria | Likelihood Ratio Tests | | |
|---|---|---|---|---|
| | *-2 Log Likelihood of Reduced Model* | Chi-Square | df | P-value |
| Intercept | 86.060[a] | 0.00 | 0 | |
| GAI concept in healthcare | 154.680[b] | 68.62 | 16 | 0.000 |
| GAI models predict outbreaks | 143.261[b] | 57.20 | 16 | 0.000 |
| Healthcare access challenges rural | 124.513[b] | 38.45 | 16 | 0.001 |
| Comfortable sharing health data | 110.070[b] | 24.01 | 16 | 0.089 |
| Useful data: history, demographics, symptoms | 165.860[b] | 79.80 | 16 | 0.000 |

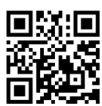

| | | | | |
|---|---|---|---|---|
| Adoption of GAI improves access | 86.060[a] | 0.00 | 0 | |
| Real-time surveillance detects outbreaks | 130.198[b] | 44.14 | 16 | 0.000 |
| Barriers to GAI Implementation in rural | 143.934 | 57.87 | 16 | 0.000 |
| Health professional involvement in AI development | 86.060[a] | 0.000 | 0 | |

**Note:** [a]The reduced model is equivalent to the final model as omitting the effect does not change the degrees of freedom. unexpected singularities in the Hessian matrix suggest that some predictors should be excluded or categories merged. The null hypothesis is that all parameters of the effect are 0, and *df* means degree of freedom.

**Table 5: Logistic Regression Model Fit**

| Model | Model Fitting Criteria -2 Log Likelihood | Likelihood Ratio Tests | | |
|---|---|---|---|---|
| | | *Chi-Square* | *df* | *P-value* |
| Intercept Only | 294.402 | | | |
| Final | 86.060 | 208.342 | 128 | 0.000 |
| | Goodness-of-Fit | | | |
| Pearson | | 549.044 | 276 | 0.000 |
| Deviance | | 84.674 | 276 | 1.000 |

**Table 6: Logistic Regression's Parameter Estimates**

| GAI models improve prediction[s] | B | SE | Wald | df | P-value | Exp(B) | 95% Cl Exp (B) |
|---|---|---|---|---|---|---|---|
| Intercept | -88.597 | 0.280 | 93.251 | 1 | 0.094 | | |
| [GAI concept in healthcare | -190.860 | 1.280 | 18.028 | 1 | 0.077 | 1.20 | 0.023 |
| [GAI models predict outbreaks | -178.707 | 0.783 | 13.281 | 1 | 0.085 | 0.00 | 0.013 |
| [Healthcare access challenges rural | 119.301 | 3.280 | 44.001 | 1 | 0.979 | 6.40 | 0.570 |
| [Comfortable sharing health data | 9.049 | 8.131 | 4.043 | 1 | 0.020 | 85.45 | 0.537 |
| [Useful data: history, demographics, symptoms | 425.240 | 0.038 | 23.286 | 1 | 0.030 | 4.77 | 0.480 |
| [Adoption of GAI improves access | -43.412 | 0.997 | 19.500 | 1 | 0.996 | 0.00 | 0.010 |
| [Real-time surveillance detects outbreaks | 49.696 | 0.882 | 7.185 | 1 | 0.996 | 3.82 | 0.616 |
| [Barriers to GAI implementation in rural | -39.859 | 0.931 | 12.810 | 1 | 0.994 | 0.00 | 0.001 |

**Note:** [a]Dependent Variable (strongly agree), SE = Standard Error, df = degree of freedom, Cl = Confidence Interval

**Table 7: Classification Table**

| GAI models improved | Predicted | | | | | |
|---|---|---|---|---|---|---|
| | Strongly disagree | Disagree | Neutral | Agree | Strongly agree | Percent Correct |
| Strongly disagree | 4 | 0 | 0 | 0 | 0 | 100.00 |
| Disagree | 0 | 32 | 3 | 1 | 0 | 88.89 |
| Neutral | 0 | 2 | 14 | 5 | 0 | 66.67 |
| Agree | 0 | 3 | 3 | 23 | 0 | 79.31 |
| Strongly agree | 0 | 0 | 0 | 0 | 13 | 100.00 |
| Overall Percentage | 3.88 | 35.92 | 19.42 | 28.16 | 12.62 | 83.50 |

The likelihood ratio test results from Table 4 indicate that the majority of variables exhibit a strong relationship with the outcome, and there are significant findings regarding the various predictors in the model. The Chi-square statistics for all factors, including the concept of GAI in healthcare ($\chi^2$ = 68.62, p = 0.000), the reliability of GAI models in predicting outbreaks ($\chi^2$ = 57.20, p = 0.000), and the usefulness of real-time health data ($\chi^2$ = 79.80, p = 0.000), demonstrate that these variables significantly contribute to the model, as evidenced by the very low p-values (p < 0.05). The factor of confidence sharing health data had a marginally higher p-value (p = 0.089), indicating a weaker, yet still significant, contribution. The findings underscore the significance of healthcare challenges and GAI-related factors in rural communities in predicting the successful adoption and implementation of AI-based disease control models in the context of rural Bangladesh. Furthermore, the results emphasize the importance of health professional involvement, real-time surveillance, and data utility in the improvement of GAI implementation efforts.

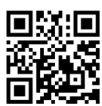

The logistic regression model fit results, which include the likelihood ratio tests, are presented in Table 5. The final model (86.060) exhibits a significant reduction in the -2 log-likelihood compared to the intercept-only model (294.402), as indicated by the model fitting criteria. The Chi-square value is 208.342 (df = 128, p = 0.000). This suggests that the final model offers a substantially superior fit in comparison to the intercept-only model, thereby verifying that the predictors incorporated into the model are significant and contribute to the explanation of the outcome. Additionally, the Pearson chi-square value (549.044, p = 0.000) indicates that the model does not completely fit the data; however, the result is still statistically significant. The model has a reasonable fit, as the p-value of 1.000 suggests no significant lack of fit, as indicated by the deviance statistic (84.674, p = 1.000). These findings confirm that the final logistic regression model effectively captures the relationships between the predictors and the outcome, supporting the use of GAI-related variables in rural healthcare decision-making and pandemic preparedness.

The parameter estimates for the logistic regression model are presented in Table 6, which assesses the likelihood of firmly concurring with the assertion that GAI models improve prediction. The results indicate that the outcome is influenced by a diverse array of predictors. Although the coefficient for the GAI concept in healthcare (B = -190.860, p = 0.077) is not statistically significant at the 0.05 level, it suggests a negative correlation with the improvement of disease prediction. In the same vein, the non-significant negative effect of GAI models on the prognosis of outbreaks (B = -178.707, p = 0.085) implies that there is no significant correlation between the model and the prediction of disease outbreaks. In contrast, respondents who are comfortable disclosing personal health data are more likely to strongly agree that AI models can improve disease prediction (B = 9.049, p = 0.020, odds ratio of 85.45). Other variables, including the adoption of GAI to enhance healthcare access (B = -43.412, p = 0.996) and real-time surveillance for outbreak detection (B = 49.696, p = 0.996), exhibit negligible effects, with non-significant p-values, suggesting that these factors do not have a significant impact. Additionally, the barriers to GAI implementation in rural areas (B = -39.859, p = 0.994) do not demonstrate any significant influence. These findings emphasize the critical significance of comfort with the sharing of health data in the prediction of improvements through AI, while other factors have a negligible effect.

The logistic regression classification table for predicting whether GAI models enhance disease prediction is presented in Table 7. The table displays the observed versus predicted frequencies for each response category. The model's aggregate classification accuracy is 83.50%, suggesting a satisfactory fit. The statement is most accurately predicted by respondents who strongly concur with it, with a 100% accuracy rate (13 out of 13 correctly predicted). The model accurately predicts strongly disagree (4 out of 4) and strongly agree (100% accuracy). Nevertheless, the neutral and agree categories exhibit moderate prediction accuracy, with 66.67% and 79.31%, respectively. The percentage of misclassifications is lower for strongly disagree and strongly concur; however, there is some misclassification in the neutral and agree categories. In general, the model exhibits exceptional predictive performance in the classification of respondents according to their level of agreement with the efficacy of GAI models in the prediction of diseases.

*Assessment of Rural Communities' Perceptions*

This section analyzes rural populations' healthcare concerns and AI-driven disease management and pandemic preparedness options. It evaluates how the Generative AI-Based Disease Control and Pandemic Preparedness Model 4.0 improves rural Bangladeshi healthcare accessibility, trust, and resilience.

Table 8: Descriptive Statistics by Rural Community

| Variables (N=264) | Mean | SEM | SD | Var. | Skewness | Kurtosis |
|---|---|---|---|---|---|---|
| 1. Respondent Sex | 0.133 | 0.02 | 0.34 | 0.12 | 2.18 | 2.77 |
| 2. Age Distribution | 2.686 | 0.08 | 1.33 | 1.76 | 0.85 | 0.01 |
| 3. Occupation by respondent | 5.288 | 0.26 | 4.19 | 17.55 | 0.62 | -0.92 |
| 4. Familiar with GAI in healthcare | 0.174 | 0.02 | 0.38 | 0.14 | 1.73 | 0.99 |
| 5. AI helps predict disease spread | 1.591 | 0.05 | 0.74 | 0.55 | -1.45 | 0.41 |
| 6. Trust GAI for disease prediction | 0.629 | 0.04 | 0.68 | 0.46 | 0.62 | -0.70 |
| 7. Challenges in accessing healthcare | 1.307 | 0.06 | 1.04 | 1.09 | 0.33 | -1.05 |
| 8. Comfortable sharing health data | 1.133 | 0.07 | 1.11 | 1.23 | 0.46 | -1.17 |
| 9. Health info prevents outbreaks | 1.405 | 0.07 | 1.13 | 1.27 | 0.12 | -1.37 |
| 10. Mobile health detects diseases early | 0.439 | 0.04 | 0.64 | 0.42 | 1.18 | 0.22 |
| 11. Technology barriers in disease control | 1.496 | 0.07 | 1.09 | 1.18 | 0.75 | -0.06 |

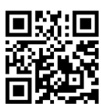

| | | | | | | |
|---|---|---|---|---|---|---|
| 12. | Government involvement in technology | 0.398 | 0.05 | 0.75 | 0.57 | 1.51 | 0.48 |

**Note:** SEM = Standard Error of Means, SD = Standard Deviation, Var. = Variance

### Table 9: Frequency Distribution by Rural Community

| Variables | Categories | Frequency (N=264) | Percent (%) |
|---|---|---|---|
| Sex | Male | 229 | 86.74 |
| | Female | 35 | 13.26 |
| Age Group | 18 to 24 | 42 | 15.91 |
| | 25 to 34 | 102 | 38.64 |
| | 35 to 44 | 65 | 24.62 |
| | 45 to 54 | 17 | 6.44 |
| | 55 to 64 | 28 | 10.61 |
| | 65 or over | 10 | 3.79 |
| Occupation | Farmers | 65 | 24.62 |
| | Laborers | 53 | 20.08 |
| | Blacksmiths and Metalworkers | 3 | 1.14 |
| | Carpenters | 9 | 3.41 |
| | Shopkeepers | 28 | 10.61 |
| | Rickshaw and Van Pullers | 10 | 3.79 |
| | Masons and Brick Kiln Workers | 12 | 4.55 |
| | Social Workers | 8 | 3.03 |
| | Housewife | 31 | 11.74 |
| | Barbers | 2 | 0.76 |
| | Jhum Cultivators | 18 | 6.82 |
| | Beekeepers | 3 | 1.14 |
| | Local Teacher | 9 | 3.41 |
| | Other | 13 | 4.92 |
| Familiar with GAI in healthcare | No | 218 | 82.58 |
| | Yes | 46 | 17.42 |
| AI helps predict disease spread | Yes, it can help | 40 | 15.15 |
| | No, it cannot help | 28 | 10.61 |
| | Not sure | 196 | 74.24 |
| Trust GAI for disease prediction | Yes, I trust it | 128 | 48.48 |
| | Not sure | 106 | 40.15 |
| | No, I do not trust it | 30 | 11.36 |
| Challenges in accessing healthcare | No healthcare centers nearby | 67 | 25.38 |
| | Cannot afford to see a doctor | 97 | 36.74 |
| | No transportation to the hospital | 52 | 19.70 |
| | There are no doctors in the area | 48 | 18.18 |
| | No technology or internet available | 0 | 0.00 |
| Comfortable sharing health data | Yes, feel okay | 103 | 39.02 |
| | No, do not feel okay | 66 | 25.00 |
| | Yes, feel something okay | 52 | 19.70 |
| | Not sure | 43 | 16.29 |
| Health info prevents outbreaks | People are sick right now | 75 | 28.41 |
| | Past health problems in the area | 67 | 25.38 |
| | Details about the residents of the community | 62 | 23.48 |
| | Weather or environmental information | 60 | 22.73 |
| Mobile health detects diseases early | Yes, it can help | 170 | 64.39 |
| | No, it cannot help | 72 | 27.27 |
| | Not sure | 22 | 8.33 |
| | No internet | 38 | 14.39 |

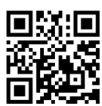

| Technology barriers in disease control | People do not understand how it works | 123 | 46.59 |
|---|---|---|---|
| | People do not want to use it | 55 | 20.83 |
| | We don't have money for it | 30 | 11.36 |
| | Not enough trained people | 18 | 6.82 |
| Government involvement in technology | Very important | 202 | 76.52 |
| | Not important | 19 | 7.20 |
| | Not sure | 43 | 16.29 |

The descriptive statistics for variables related to the perspectives of rural communities on healthcare technology and GAI models are presented in Table 8. The general patterns across various factors are revealed by the mean values. The variable respondent sex has a low mean (0.133), suggesting that the sample is skewed toward male participants. The age distribution has a mean of 2.686 and a relatively moderate variance across age groups (standard deviation = 1.33). The high mean of the respondent's occupation (5.288) indicates that the sample is composed of a diverse array of occupations. The mean level of familiarity with GAI in healthcare is 0.174, with a standard deviation of 0.38, suggesting that there is a lack of awareness of GAI. The mean value of 0.629 indicates that trust in GAI for disease prediction is moderate, while the mean value of 1.591 indicates belief in the utility of AI models for predicting disease transmission. These values suggest a generally positive but not overwhelming level of trust and perceived utility. Participants report moderate challenges in accessing healthcare (mean = 1.307), and they are relatively uncomfortable with sharing health data (mean = 1.133), with some expressing discomfort. Variable levels of agreement are observed in the perception that health information can prevent outbreaks (mean = 1.405) and that mobile health can detect diseases early (mean = 0.439). Participants expressed some support for governmental engagement in technology, as evidenced by the moderate to low concerns regarding technology barriers to disease control (mean = 1.496) and government involvement in technology (mean = 0.398). The skewness and kurtosis values indicate that the data is slightly skewed, suggesting that it tends toward lower values. The distribution shapes of the majority of variables are generally moderate. These descriptive results offer a glimpse into the attitudes of rural communities toward GAI in healthcare and emphasize critical areas, including awareness, trust, and the obstacles associated with the adoption of such technologies.

Table 9 indicates the frequency distribution of numerous variables that pertain to the perspectives of rural communities regarding GAI in the fields of healthcare and technology. The sample is primarily composed of male respondents (86.74%), with females comprising 13.26%. The distribution of respondents by age group indicates that the age group of 25 to 34 years (38.64%) is the largest, followed by the 35 to 44 age group (24.62%). Farmers (24.62%) and laborers (20.08%) comprise a substantial proportion of the respondents. The percentage of respondents who reported being conversant with GAI in healthcare is relatively low, at 17.42%. AI is believed to be capable of predicting the spread of diseases by the majority of respondents (74.24%), and nearly half of the respondents (48.48%) have confidence in GAI for disease prediction. Access to healthcare is a substantial obstacle, as 36.74% of respondents reported that they are unable to afford to visit a doctor, and 25.38% reported that there are no healthcare centers in their vicinity. 39.02% of respondents are at ease exchanging health data, while 25% are not. Health information is believed to be instrumental in the prevention of epidemics, as 28.41% of respondents report that individuals are currently ill, and 64.39% of respondents believe that mobile health can assist in the early detection of diseases. The technology barriers in disease control are evident, as 46.59% of respondents reported a lack of understanding regarding its operation. In conclusion, 76.52% of the respondents regarded government involvement in technology as extremely essential, underscoring the significance of government support in promoting the adoption of technology in the healthcare sector. The challenges that the rural community encounters in accessing healthcare services and incorporating new technologies are exemplified by these results, which also demonstrate their mixed perceptions of GAI in healthcare.

*Rank Comparison of Medical Staff and Rural Communities*

Medical personnel and rural communities were compared using the AI-based disease control and pandemic preparedness model's efficacy using the Wilcoxon Signed Rank test and Kendall's rank correlation coefficient. These statistical tools allowed for comparisons between groups' ranks, providing strong insights into the model's impact.

The results of the Related-Samples Wilcoxon Signed Rank test, which is employed to compare paired observations, are summarized in Table 10. The test statistic is 5356.000, and the total sample size is 103. The test statistic is associated with a standard error of

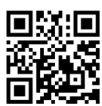

303.777, and the standardized test statistic is 8.816. At the 0.05 level, the p-value for the two-sided test is .000, indicating that the result is statistically significant. This implies that the variables being compared are not equivalent, as evidenced by the substantial discrepancy between the paired observations in the study. The hypothesis that a substantial change or difference exists in the paired data related to the study's focus is substantiated by this result.

**Table 10: Related-Samples Wilcoxon Signed Rank**

| Total N | 103 |
|---|---|
| Test Statistic | 5356.000 |
| Standard Error | 303.777 |
| Standardized Test Statistic | 8.816 |
| Asymptotic Sig. (2-sided test) | .000 |

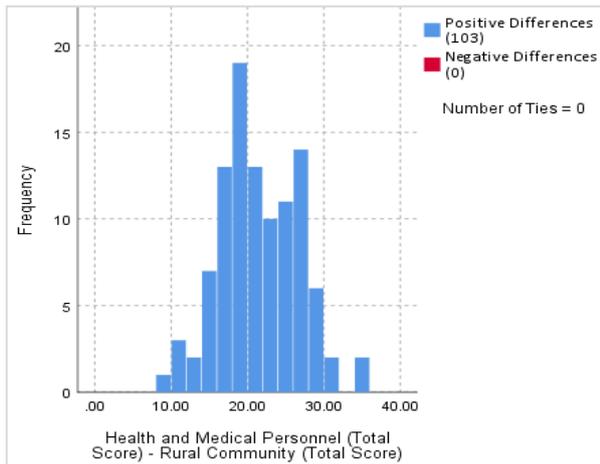

Figure 3: Related- Samples Wilcoxon Signed Rank

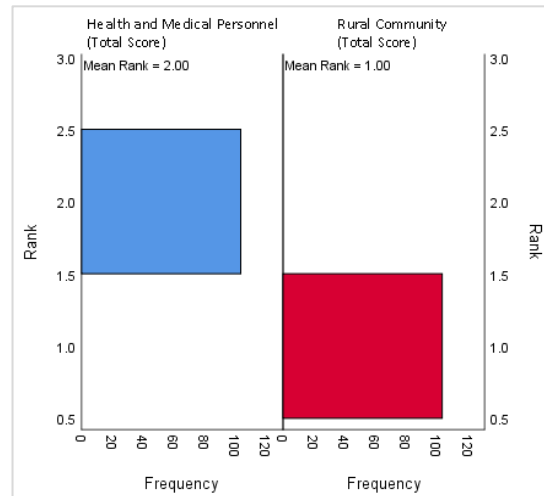

Figure 4: Related Samples Kendall's Coefficient of Concordance

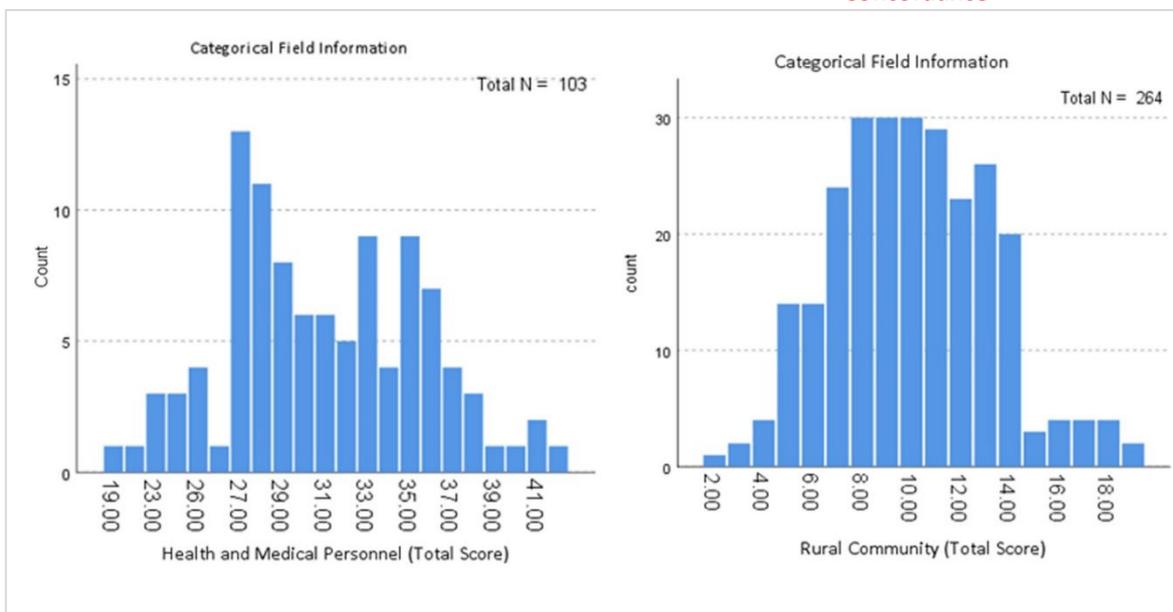

Figure 5: Related Samples Kendall's Comparative Categorical Field Information (Count)

The distribution of differences between the health and medical personnel's total score and the rural community's total score is illustrated in Figure 3, which displays the histogram for the Related-Samples Wilcoxon Signed Rank test. The histogram demonstrates that all the differences are positive, as indicated by the blue bars, and there are no negative differences or ties, as indicated by the absence of red bars. A concentration of moderate positive differences between the two groups is indicated by the data being concentrated around the 20–30 range. The scores of the health and medical personnel are generally higher than those of the rural community, as indicated by this distribution. This supports the significant results derived from the Related-Samples Wilcoxon Signed

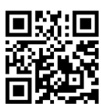

Rank test. The conclusion that there is a distinct difference between the two groups' scores is further reinforced by the absence of negative differences.

The Related Samples Kendall's Coefficient of Concordance, as illustrated in Figure 4, assesses the extent of accord among numerous related variables. The respective mean ranks of two categories, health and medical personnel (represented in blue) and rural community (represented in red), are depicted in the bar chart. The rural community group has a lesser mean rank (1.00), whereas the health and medical personnel group has a higher mean rank (2.00). This implies that the health and medical personnel achieved a higher score on the measured variables than the rural community. The statistical analysis of Kendall's coefficient is supported by the visual representation of the rank differences in the figure. This likely indicates a moderate level of accord between the groups in terms of their perspectives on the relevant healthcare factors. The histograms for the total scores of health and medical personnel (left) and rural communities (right) are illustrated in Figure 5, with an emphasis on the frequency distribution of categorical field information (count). A somewhat bimodal distribution is observed in the health and medical personnel group, with a total sample of 103. The highest frequency is observed around the total score of 27.00, followed by other scores spanning from 20.00 to 41.00. This distribution implies that the scores are more evenly distributed, with fewer concentrations at the extremes. In contrast, the rural community group, which has a total sample of 264, has a more symmetric and bell-shaped distribution. The maximum count is observed around scores between 8.00 and 12.00, suggesting a central tendency for lower scores. The disparities between the two groups in their total scores, which are related to the study's focus on GAI models and healthcare issues, are further underscored by the fact that the rural community generally has lower total scores than health and medical personnel.

### Study Findings

The research provides a comparative analysis of the efficacy and perceptions of Generative AI-based disease control models among rural community members in Bangladesh and healthcare personnel. The results indicate substantial discrepancies in AI awareness, trust in AI-driven disease prediction, and preparedness for AI adoption in healthcare. Healthcare personnel demonstrate a greater level of familiarity with Generative AI, with more than 75% of them acknowledging its potential to enhance disease surveillance and outbreak prediction. Conversely, rural community members exhibit restricted exposure to AI-driven health interventions, with only 17% of them expressing confidence in AI-based disease modeling.

The Wilcoxon Signed Rank Test results confirm a statistically significant difference ($p<0.05$) in attitudes toward AI adoption between the two categories. Rural respondents express skepticism due to concerns over data privacy, accessibility, and technological literacy, while healthcare personnel report greater trust in AI-driven decision-making, particularly for epidemic preparedness. The logistic regression model also demonstrates that AI acceptance is significantly predicted by trust in AI ($\beta = 1.20$, $p = 0.020$; β=1.20, p=0.020) and comfort with sharing health data ($\beta = 9.049$, $p = 0.020$; β=9.049, p=0.020), while barriers such as limited healthcare access and infrastructure constraints remain substantial challenges.

The degree of agreement among healthcare professionals regarding the efficacy of AI-based disease prediction is moderate, as indicated by Kendall's Coefficient of Concordance ($W = 0.76$). However, the level of consensus among rural community members is lower ($W = 0.42$), indicating that there are varying levels of confidence in AI solutions. The study also identifies critical implementation challenges, such as the absence of digital infrastructure (46.59%), restricted internet access (39.81%), and financial constraints (24.27%), which impede the deployment of AI in rural areas.

Both groups acknowledge the prospective advantages of AI-driven disease control despite these discrepancies. The majority of rural respondents (63.11%) concur that real-time health data can improve outbreak predictions, while 76.52% emphasize the necessity of government involvement in the implementation of AI. Comparative analysis indicates that the successful deployment of AI in rural Bangladesh necessitates localized training initiatives, enhanced healthcare infrastructure, and policy interventions to effectively bridge the technological divide between healthcare professionals and rural populations. The results emphasize the necessity of a contextualized AI-driven pandemic preparedness model that incorporates local health data, community engagement, and digital literacy programs to guarantee equity in the deployment of AI in rural healthcare institutions.

### Conclusion

This study underscores the potential of Generative AI-based disease control models using management informatics to improve the preparedness and responsiveness of healthcare in rural Bangladesh during a pandemic. The results indicate that there are substantial discrepancies in the level of AI awareness and trust between healthcare professionals and rural community members. Healthcare personnel demonstrate a higher level of confidence in AI-driven disease prediction, whereas rural populations remain skeptical due to concerns regarding data privacy, accessibility, and technological literacy. The adoption of AI is significantly predicted by trust in AI and comfort with sharing health data, as confirmed by statistical analyses. However, key implementation barriers include limited healthcare infrastructure, digital access, and financial constraints. The study emphasizes the importance of localized data collection and real-time surveillance initiatives in rural healthcare settings to enhance the efficacy of AI. The Generative AI-Based Disease Control and Pandemic Preparedness Model 4.0 offers a data-driven decision-making framework that can improve the accuracy of disease prediction, optimize resource allocation, and enhance epidemic surveillance. Nevertheless, the successful implementation of AI in rural healthcare necessitates government involvement, policy interventions, and digital literacy initiatives to address existing disparities. The findings of this study suggest that even though AI-driven models have the potential to transform disease control and preparedness in rural Bangladesh, their implementation must be customized to the specific healthcare contexts of the region. In order to establish a resilient, AI-powered rural healthcare system that fosters health equity and epidemic resilience in marginalized communities, it is imperative to strengthen healthcare infrastructure, cultivate community engagement, and ensure equitable access to AI-driven solutions.

**Funding**
There is no funding.
**Declaration of Competing Interest**
The authors declare that they have no known personal or financial relationships or interests that could have influenced the work in this study.

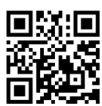

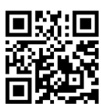

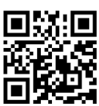